\documentclass[eqsecnum,aps,epsf,groupedaddress]{revtex4}
\usepackage{latexsym}
\usepackage{amssymb} 
\usepackage{amsmath}
\begin{document}

\title{BPS pp-wave brane cosmological solutions in string theory}

\author{Makoto Tanabe
\footnote{e-mail address:tanabe@gravity.phys.waseda.ac.jp}}
\affiliation{Department of Physics Waseda University, 
Okubo 3-4-1, Shinjuku Tokyo 169 Japan}
\author{Shuntaro Mizuno
\footnote{e-mail address:shuntaro@gravity.phys.waseda.ac.jp}}
\affiliation{Department of Physics Waseda University, 
Okubo 3-4-1, Shinjuku Tokyo 169 Japan}

\date{\today}

\begin{abstract}

We construct time dependent  BPS pp-wave brane solutions 
in the context of M-theory and type II supergravity. 
It is found that N-brane solutions we considered 
satisfy the crossing rule 
as S-brane solutions but 1/8 supersymmetry remains. 
By applying them to the cosmological setting,
inflationary solutions are obtained.
During this inflation, the size of the extradimensions becomes
smaller than our four-dimensional spacetime dynamically.
We also discuss the mechanism for terminating this inflation
and recovering the hot big-bang universe.
\end{abstract}

\maketitle

\section{Introduction}

With the strong support of observational results,
the standard big-bang cosmology supplemented by the inflationary scenario
has come to rise in status \cite{inflation}. 
What is necessary next, hopefully,
is to explain the corresponding cosmological solutions in the context
of a fundamental theory such as string theory. Regardless of the conceptual
worries in the study of dS or eternally accelerating universe 
quantum gravity \cite{Banks:2000fe, Strominger:2001pn}, 
one possibility to explain
de Sitter vaccua \cite{Kachru:2003aw}
as well as inflationary solutions \cite{Kachru:2003sx} has
recently been proposed in the context of string theory.
It is known that the no-go theorem of \cite{Maldacena:2000mw}
guarantees that such solutions cannot be obtained in string or M theory
by using only the lowest-order terms in the 10D or 11D supergravity
action. The crucial elements to invalidating the no-go theorem
are the D-branes and fluxes supported by form fields 
in warped backgrounds and allow one to find highly warped 
compactification such as \cite{Giddings:2001yu}.

D-branes \cite{Polchinski:1995mt} have played a very important
role not only in cosmology but also in our understanding of non-perturbative
aspects of string or M theory and the AdS/CFT correspondence 
\cite{Maldacena:1997re}. As is well known, D-branes can be described
as hypersurfaces where open strings can end, which is achieved by imposing 
Dirichlet boundary conditions along transverse spacelike directions
in the string world-sheet action, in perturbative string theory
at weak string coupling. The more general situations in with D-branes
can be understood as intersecting ones, with which explicit string 
theory compactifications have appeared in \cite{Blumenhagen:2000wh}.
What is interesting about the intersecting branes phenomenologically
is pointed out in \cite{Berkooz:1996km};
when D-branes intersect at non-vanishing angles, open string
stretched out between them gives rise to chiral fermions 
living at the intersection.

Naturally, in string perturbation theory, one can also consider open 
strings obeying Dirichlet boundary conditions along time-like or
null directions. These are space-like or null analogs of D-branes
and are called S-branes \cite{Gutperle:2002ai} or 
N-branes \cite{Kogan:2001nn}, respectively. Until now, even though 
time-dependent brane solutions can be considered in both cases,
relatively, S-brane solutions acquire much more attention because
of their possible connection with rolling tachyon and dS/CFT
correspondence \cite{Strominger:2001pn, Gutperle:2002ai,
Chen:2002yq, Kruczenski:2002ap, Roy:2002ik, Bhattacharya:2003sh,
Hashimoto:2002sk, Jones:2004rg, Tasinato:2004dy} 
as well as inflationary
solutions \cite{Ohta:2003pu} (see also Refs.~\cite{Behrndt:1994ev,
Hull:1998vg} for related solutions). As in D-brane cases, intersecting
S-brane solutions can be also obtained \cite{Deger:2002ie, 
Ivashchuk:1997pk, Ohta:2003uw}. On the other hand, N-brane solutions are 
also interesting from the viewpoint of closed/open string correspondence
and stringy explanation of the black holes as discussed in 
Ref.~\cite{Kogan:2001nn}, where such solutions were discussed in the
string worldsheet picture. Recently, some class of explicit intersecting 
N-brane solutions in supergravity have been obtained \cite{Ohta:2003zh}
in which the intersection rules for the way the solutions can intersect with
each other is given based on the method of 
\cite{Ohta:2003uw, Ohta:1997gw}. The main purpose of this paper is
to construct on other class of N-brane cosmological solutions 
that are reminiscent of a stringy set-up.

In order to start with the string theory background, we adopt pp-wave 
spacetime, which yields exact classical 
backgrounds for string theory for some time, with all $\alpha'$ corrections 
vanishing \cite{Amati:1988sa,Horowitz:1989bv}. It is also worth noting
that these backgrounds are exactly solvable in the light cone gauge
\cite{Metsaev:1999gz, Metsaev:2001bj, Metsaev:2002re}. 
The outline of our paper is as follows.
In Section II after presenting the low-energy effective action 
corresponding to superstring theory in any spacetime dimension,
we obtain the basic equations. 
In Section III we assume the pp-wave background and 
the gauge field strength, and the 1/8 supersymmetric BPS solutions 
can be obtained. We also found that the crossing rule and the 
harmonic rule are satisfied under the metric ansatz.  
In Section IV we indicate the solutions in M or string theory, 
and we show the relations between the S-brane solutions 
\cite{Ohta:2003uw}. 
In Section V we find specific solutions corresponding to the 
inflationary solutions in four dimensions. These solutions 
provide dynamical compactification naturally.
Section VI discusses the end of the inflation via supersymmetry breaking.

\section{Model and Basic Equations}

Consider the following general action for gravity coupled to 
a dilaton $\phi$ and $m$ different $n_A$- form field strengths:

\begin{equation}
S=\frac{1}{16\pi G_D}\int d^Dx\sqrt{-g}
\left[R-\frac{1}{2}(\nabla\varphi)^2
-\sum _{A=1}^{m}\frac{1}{2\cdot n_A!}
e^{a_A\varphi}F_{n_A}^2\right],
\label{action}
\end{equation}
where $R$ is the Ricci scalar with respect to the metric $g_{\mu\nu}$, 
$\varphi$ is the dilaton,
$F_{n_A}$ is field strengths 
of arbitrary form degree $n_A\leq D/2$, and $a_A$ 
is the coupling between the dilaton and the form field.
This action describes the bosonic part of $D=11$ or $D=10$
supergravities.
In addition to this, in general, 
there may be Chern-Simons terms in the action.
However, since  they are irrelevant in our following 
solutions, we omit them by assuming they are frozen.

The equations of motion 
corresponding to the Einstein equation, 
the equation of motion for the dilaton,
and the Maxwell equation can be written 
in the following forms, respectively: 
\begin{eqnarray}
&&R_{\mu\nu}=\frac{1}{2}\nabla _\mu\varphi\nabla _\nu
\varphi+\sum _A\Theta _{A\mu\nu},\nonumber\\
&&\nabla^2\varphi =\sum _A\frac{a_A}{2\cdot 
n_A!}e^{a_A\varphi}F_{n_A}^2,\nonumber\\
&&\partial _{\mu _1}(\sqrt{-g}e^{a_A\varphi}F^{\mu _1
\cdots\mu _{n_A}})=0,\label{eqn:eom}
\end{eqnarray}
where $\Theta _{A\mu\nu}$ is the 
stress-energy tensor corresponding to the $n_A$-form field, 
given by
\begin{eqnarray}
\Theta _{A\mu\nu}=\frac{1}{2\cdot n_A!}
e^{a_A\varphi}\left[n_A{F_{\mu}}^{\rho\cdots
\sigma}F_{\nu\rho\cdots\sigma}-\frac{n_A-1}{D-2}
F_{n_A}^2g_{\mu\nu}\right],
\end{eqnarray}
and $\nabla^2$ is a $D$-dimensional Laplacian 
with respect to $g_{\mu\nu}$.

The Bianchi identities for the $n_A$-form
serve as the constraint equation, which is
\begin{equation}
\partial _{[\mu}F_{\mu _1\cdots\mu _{n_A}]}=0,
\end{equation}
as they are the field strength of $(n_A-1)$-form potentials.

\section{PP-wave Solutions}

In this section we find the pp-wave background solution
for the basic equaions (\ref{eqn:eom}).  
We take the following metric for our system:

\begin{equation}
ds^2=2e^{2\xi}du(dv+fdu)+\sum _{i=1}^{d-1}e^{2\eta}dx_i^2
+\sum _{\alpha=2}^pe^{2\zeta _\alpha}dy_\alpha^2,
\label{metric_ansatz}
\end{equation}
where we have used the light-cone coordinate $u=-(t-y_1)/\sqrt{2}$ 
and $v=(t+y_1)/\sqrt{2}$. Furthermore, as for the number of spacetime
dimensions, we separate them into $D=d+p$, the coordinates $y_\alpha$, 
$\alpha=2,...,p$ parametrize the $(p-1)$-dimensional world-volume 
directions, and the remaining coordinates of the $(d+1)$-dimensional
spacetime are coordinates on $(d-1)$-dimensional flat spaces, $u$
and $v$. Since we are interested in time-dependent solutions,
all the functions appearing in the metrics as well as the dilaton
$\varphi$ are assumed to depend only on the light-cone 
coordinate $u$ and $v$. These solutions about the light cone 
coordinate are named ``null-branes"(N-branes) and 
the ${\rm N}{q_A}$-brane whose world-volume is $(q_A + 1)$-
dimensional, tangential to $u$, and $q_A$-
spacelike directions. (We assume that these spatial coordinates
correspond to some of $\{y_1,\cdots ,y_p\}$ in our solutions.)

As for the form fields, the most general ones consistent with the
field equations and Bianchi identities should be taken.
For this purpose, we assume an electrically charged ${\rm N}{q_A}$-
brane whose value is given by 
\begin{eqnarray}
F_{n_A}=d\phi_{q_A}=\partial_vE_{q_A}du\wedge dv\wedge
dy_2\wedge\cdots\wedge dy_{q_A+1}, 
\label{elec_ansatz}
\end{eqnarray}
where $n_A=q_A+2$ and $E_{q_A}$ is a function of $u$ 
and $v$. We can also discuss the magnetic case with a 
N$\tilde{q_A}$-brane, which is obtained from a dual 
transformation of the electrically charged 
N$q_A$-brane as 
\begin{eqnarray}
\tilde{F}_{n_A}&=&e^{-a_A\varphi}\partial_vE_{\tilde{q}_A}
e^{-2\xi-\sum_\alpha\zeta_\alpha}{\epsilon^{uv\alpha_2 
\cdots\alpha_{\tilde{q}_A+1}}}_{1\cdots (d-1)
\alpha_{\tilde{q}_A+2}
\cdots \alpha_p}\nonumber\\
&&\quad dx^1\wedge\cdots\wedge dx^{d-1}\wedge 
dy^{\alpha_{\tilde{q}_A+2}}\wedge\cdots\wedge dy^{\alpha_p}.
\label{magn_ansatz}
\end{eqnarray}

In order to treat both types of the null branes 
simultaneously, we define the new function as 
\begin{eqnarray}
C_A \equiv \partial_vE_Ae^{\epsilon_Aa_A\varphi-2\xi-\sum\zeta_\alpha},
\end{eqnarray}
where 
\begin{eqnarray}
\epsilon_A=\left\{\begin{array}{ll}
+1 & \mbox{electric field} \\
-1 & \mbox{magnetic field.}\end{array}\right.
\end{eqnarray}

\subsection{Bosonic part of the solutions and crossing rule}
Here we solve the bosonic part basic equations under the ansatzs
(\ref{metric_ansatz}), (\ref{elec_ansatz}), and (\ref{magn_ansatz})
as generally as possible and derive the general intersecting rules for
N-branes.

By choosing the gauge satisfying
\begin{eqnarray}
(d-1)\eta+\sum_{\alpha=2}^{p}\zeta_\alpha=0,
\label{gauge_cond}
\end{eqnarray}
the field equations can be expressed
as follows: 
\begin{eqnarray}
&&(d-1)(\partial_u\eta-f\partial_v\eta)^2
+\sum_{\alpha}(\partial_u\zeta_\alpha-f\partial_v\zeta_\alpha)^2
=-\frac{1}{2}(\partial_u\varphi-f\partial_v\varphi)^2
-\sum_A2fC_A\partial_vE_A,\label{eqn:Ruu}\\
&&2\partial_v(\partial_u\xi-f\partial_v\xi)-\partial_v^2f
+(d-1)(\partial_u\eta-f\partial_v\eta)\partial_v\eta
+\sum_\alpha(\partial_u\zeta_\alpha-f\partial_v\zeta_\alpha)
\partial_v\zeta_\alpha\nonumber\\
&&\quad=-\frac{1}{2}(\partial_u\varphi-f\partial_v\varphi)
\partial_v\varphi+\sum_A\frac{D-q_A-3}{2(D-2)}2C_A
\partial_vE_A,\label{eqn:Ruv}\\
&&(d-1)(\partial_v\eta)^2+\sum_\alpha(\partial_v\zeta_\alpha)^2
=-\frac{1}{2}(\partial_v\varphi)^2,\label{eqn:Rvv}\\
&&\partial_v(\partial_u\eta-f\partial_v\eta)=
-\sum_A\frac{q_A+1}{2(D-2)}C_A\partial_vE_A,\label{eqn:Rii}\\
&&\partial_v(\partial_u\zeta_\alpha-f\partial_v\zeta_\alpha)=
\sum_A\frac{\delta_{\alpha A}}{2(D-2)}C_A\partial_vE_A,\label{eqn:Ryy}\\
&&\partial_v(\partial_u\varphi-f\partial_v\varphi)=
-\sum_A\frac{1}{2}\epsilon_Aa_AC_A\partial_vE_A,\label{eqn:dilaton}\\
&&\partial_uC_A=\partial_vC_A=\partial_v(fC_A)=0,
\label{eqn:maxwell}
\end{eqnarray}
where $\partial_u$ and $\partial_v$ are partial derivatives 
$\partial/\partial u$ and $\partial/\partial v$ in the light-cone 
coordinate and $A$ denotes the kinds of N$q_A$-branes.
$\delta_{\alpha A}$ is the constant decided by 
\begin{eqnarray}
\delta_{\alpha A}=\left\{\begin{array}{ll}
D-q_A-3 & \mbox{$y_\alpha$ belonging to $q_A$-brane} \\
-(q_A+1) & \mbox{otherwise}\end{array}\right.
\end{eqnarray} 
Eqs.~(\ref{eqn:Ruu})-~(\ref{eqn:Ryy}) are the $(u,u)$, $(u,v)$, 
$(v,v)$, $(i,i)$, $(\alpha,\alpha)$ components of the Einstein equation 
in Eq.~(\ref{eqn:eom}),
respectively.
From the Maxwell equations (\ref{eqn:maxwell}),
$C_A$ is shown to be a constant and  $\partial_vf=0$, which
suggests $f=f(u)$.  

Since $C_A$ is a constant in
Eq.~(\ref{eqn:Rii}), (\ref{eqn:Ryy}), and the dilaton 
equation (\ref{eqn:dilaton}), 
\begin{eqnarray}
&&\partial_v\left[(\partial_u-f\partial_v)\eta+
\sum_A\frac{q_A+1}{2(D-2)}C_AE_A\right]=0,
\label{eqn:etaderivative}\\
&&\partial_v\left[(\partial_u-f\partial_v)\zeta_\alpha-
\sum_A\frac{\delta_{\alpha A}}{2(D-2)}C_AE_A\right]=0,\\
&&\partial_v\left[(\partial_u-f\partial_v)\varphi+
\sum_A\frac{1}{2}\epsilon_Aa_AC_AE_A\right]=0,
\end{eqnarray}
and $\partial_vf=0$ in Eq.~(\ref{eqn:Ruv}) provides
\begin{eqnarray}
\partial_v\left[(\partial_u-f\partial_v)\xi-
\sum_A\frac{D-q_A-3}{2(D-2)}C_AE_A\right]=0.
\end{eqnarray}

In this paper, we concentrate on solutions with the following conditions, 
which satisfy the above equations automatically: 
\begin{eqnarray}
&&(\partial_u-f\partial_v)\xi=\sum_A\frac{D-q_A-3}{2(D-2)}
C_AE_A,\label{eqn:difxi}\\
&&(\partial_u-f\partial_v)\eta=-\sum_A\frac{q_A+1}{2(D-2)}
C_AE_A,\label{eqn:difeta}\\
&&(\partial_u-f\partial_v)\zeta_\alpha=\sum_A
\frac{\delta_{\alpha A}}{2(D-2)}C_AE_A,\label{eqn:difzeta}\\
&&(\partial_u-f\partial_v)\varphi=-\sum_A\frac{1}{2}
\epsilon_Aa_AC_AE_A.\label{eqn:difdilaton}
\end{eqnarray}

We show these conditions are consistent with the BPS condition, 
of an extremal solution in supergravity, as we will see 
in the next subsection.
Substituting Eqs.~(\ref{eqn:difeta}), (\ref{eqn:difzeta}), and 
(\ref{eqn:difdilaton}) to Eq.~(\ref{eqn:Ruv}), we find 
\begin{eqnarray}
\sum_{A,B}\left[M_{AB}\frac{C_A}{2}+\delta_{AB}
\partial_v\left(\frac{f}{E_A}\right)\right]\frac{C_B}{2}E_AE_B=0, 
\label{eqn:difRuv}
\end{eqnarray}
where $M_{AB}$ is a constant matrix defined as 
\begin{eqnarray}
M_{AB}=(d-1)\frac{(q_A+1)(q_B+1)}{(D-2)^2}+
\frac{1}{2}
\epsilon_Aa_A\epsilon_Ba_B
\sum_{\alpha=2}^{p}\frac{\delta_{\alpha A}\delta_{\alpha B}}{(D-2)^2}.
\label{eqn:MAB}
\end{eqnarray}

Notice that until now our discussion is quite general
except for imposing on the ansatzs (\ref{metric_ansatz}), 
(\ref{elec_ansatz}) and (\ref{magn_ansatz}), and 
the gauge condition (\ref{gauge_cond}).
From Eq.~(\ref{eqn:difRuv}) an important condition is derived
if we require that the functions $E_A$ with different index
$A$ are independent, that is, 
$M_{AB}=0$ 
for $A\neq B$. Suppose that N$q_A$-brane and N$q_B$-brane 
intersect over
$\bar{q}_{AB} +1$ dimensions $(\bar{q}_{AB}<q_A, q_B)$.
A rule for the crossing dimensions, 
which is called the crossing rule of the branes is obtained as 
\begin{eqnarray}
\bar{q}_{AB}=\frac{(q_A+1)(q_B+1)}{D-2}-1-\frac{1}{2}
\epsilon_Aa_A\epsilon_Ba_B.\label{eqn:crossing}
\end{eqnarray}
This crossing rule is the same as that of a previous work
considering slightly different situations \cite{Ohta:2003zh}
and as that for the S-brane cases given by 
\cite{Deger:2002ie, Ohta:2003uw}. 

On the other hand, by considering the case
$A=B$ in Eq.~(\ref{eqn:MAB}), we have  
\begin{eqnarray}
M_{AA}=\frac{(q_A+1)(D-q_A-3)}{D-2}+\frac{1}{2}a_A^2
\equiv \frac{\Delta_A}{D-2}.
\end{eqnarray}
Therefore Eq.~(\ref{eqn:difRuv}) provides 
\begin{eqnarray}
E_A=\sqrt{\frac{2(D-2)}{\Delta_A}}\frac{f}{H_A},
\end{eqnarray}
where $H_A$ is a harmonic function on $\{u,v\}$
satisfying
\begin{eqnarray}
\partial_u\partial_vH_A=\partial_v^2H_A=0,
\label{eqn:harmonicrule}
\end{eqnarray}
which is clear from Eq.~(\ref{eqn:maxwell}).

In the following, we show examples of the solutions satisfying
the crossing rule obtained above.

In 11-dimensional supergravity, there is only a 3-form field; 
thus there is no dilaton $\varphi$.  Setting $D=11$ and $a_A=0$, 
we find that $\Delta_A=(q_A+1)(8-q_A)$. 
For the 3-form field, because $n_A=4$ the electric type field is related to 
NM2-brane, i.e., $q_A=n_A-2=2$ and $\Delta_A=18$. 
Therefore, the solutions with one electrically charged N-brane 
are then written as 

\begin{eqnarray}
ds^2_{11}&=&2 e^{2\xi}du(dv+fdu)+e^{2(\zeta_2 + \zeta_3)}
(dy_2^2+dy_3^2)+ e^{2 \eta} \sum_{i=1}^7dx_i^2,\nonumber\\
 F_4&=&-d(f/H_2)\wedge du\wedge dy_2\wedge dy_3,
 \label{sol:single_nm2}
 \end{eqnarray}
 where $H_2$ is a harmonic function depending on $u$ and $v$.
 
 On the other hand, the magnetic type field is related to the NM5-brane 
 because $\tilde{q}_A=\tilde{n}_A-2=D-n_A-2=5$, which provide
 $\Delta_A=18$. 
 Therefore, the solutions with one magnetically charged N-brane 
are given by
 \begin{eqnarray}
ds^2_{11}&=&2 e^{2\xi} du(dv+fdu)+ 
\sum_{\alpha=2}^6 e^{2 \zeta_\alpha} dy_\alpha^2+
e^{2 \eta} \sum_{i=1}^4 dx_i^2,
\nonumber\\ \ast F_4&=&H_5^2 \ast d(f/H_5),
\label{sol:single_nm5}
\end{eqnarray}
where $H_5$ is a harmonic function depending on $u$ and $v$, too. 

Of course, it is also possible to introduce the combinations of
NM2-branes and NM5-branes. In such cases, the crossing rule obtained
in Eq.~(\ref{eqn:crossing}) plays a very important role.
From the crossing rule, all the possible cases for the intersecting
dimensions are:  
\begin{eqnarray}
M2\cap M2\rightarrow \bar{q}=0,\quad
M2\cap M5\rightarrow \bar{q}=1,\quad
M5\cap M5\rightarrow \bar{q}=3.
\end{eqnarray}
Among them, 
we obtain $d=4$ the case with the BPS pp-wave solutions uniquely as follows:  
\begin{eqnarray}
ds^2_{11}&=&2 e^{2\xi}du(dv+fdu)
\sum_{\alpha=2}^7 e^{2 \zeta_\alpha} dy_\alpha^2+
e^{2 \eta} \sum_{i=1}^3 dx_i^2,
\label{sol:cross_nm2_nm5}
\end{eqnarray}
in which  the NM$2$-brane occupies the $u, y^2$ and $y^7$ 
directions and NM$5$-brane occupies the $u, y^2\ldots , y^6$ directions.
The solution given by Eq.~(\ref{sol:cross_nm2_nm5}) is especially
interesting in that it produces cosmological solutions after
the compactification, as we will mention later.

\subsection{BPS pp-wave brane solutions}

In the previous subsection, we notice only the bosonic part
based on the action (\ref{action}), and the solutions obtained are 
irrelevant to the fermionic part and supersymmetry. 
Here, since we are interested in supersymmetry and the BPS solutions,
we construct concrete BPS solutions. 
As follows, we can investigate the condition under which 
supersymmetry remains without writing down the details of 
the fermionic part. This condition can be attained by considering 
the supersymmetry transformation of the gravitino $\psi_\mu$,
which is called the Killing equation for the Killing spinor
$\zeta$ \cite{supergravity}. Even though we consider 
only the 11-dimensional case here for simplicity,
it is worth noting that this argument can be extended to the 10-dimensional 
case easily. In the 11-dimensional case, the vanishing condition of the
supersymmetry transformation of the gravitino is given as
\begin{equation}
\delta\psi_\mu=\left[\partial_\mu+\frac{1}{4}{\omega^{ab}}_\mu
\gamma_{ab}+\frac{1}{144}({e^a}_\mu{\gamma_a}^{bcdf}
-8{e^a}_\mu\delta^b_a\gamma^{cdf})F_{bcdf}\right]\zeta=0,
\label{deltapsi_mu}
\end{equation}
where $\gamma$'s are the antisymmetrized products of 
11-dimensional gamma matrices with unit strength. 
${e^a}_\mu$ is a basis of a tetrad, and a spin connection is 
${\omega^{ab}}_\mu={e^c}_\mu{\omega^{ab}}_c$, where 
${\omega^{ab}}_\mu$ is defined by  
\begin{eqnarray}
\omega_{ab\mu}\equiv\frac{1}{2}{e_a}^\nu
(\partial_\mu e_{b\nu}-\partial_\nu e_{b\mu})
-\frac{1}{2}{e_b}^\nu(\partial_\mu e_{a\nu}-
\partial_\nu e_{a\mu})-\frac{1}{2}{e_a}^\rho
{e_b}^\sigma{e^c}_\mu(\partial_\rho e_{c\sigma}
-\partial_\sigma e_{c\rho}).
\end{eqnarray}

Now we consider the $i$-th components of the Killing equations.
For the metric given by Eq.~(\ref{metric_ansatz}),
the spin connection, ${\omega^{ab}}_i$, can be written as 
\begin{eqnarray}
{\omega^{ab}}_i&=&-e^{\eta-\xi} \partial_v\eta
(\delta^a_u\delta^b_i-\delta^a_i\delta^b_u)
+e^{\eta-\xi}\sum_A\frac{q_A+1}{18}f\partial_v\ln H_A
(\delta^a_v\delta^b_i-\delta^a_i\delta^b_v), 
\label{omega_abi}
\end{eqnarray}
where we use the conditions of (\ref{eqn:etaderivative}). 
Substituting Eq.~(\ref{omega_abi}) into Eq.~(\ref{deltapsi_mu}),
we find 
the supersymmetric solutions must satisfy 
\begin{eqnarray}
\partial_v\eta=\sum_A\frac{q_A+1}{18}\partial_v\ln H_A,
\end{eqnarray}
and we can solve this equation as 
\begin{eqnarray}
\eta=\sum_A\frac{q_A+1}{18}\ln H_A.
\end{eqnarray}

The other conditions of $\delta \psi_\mu = 0$ provide similar conditions;  
then we finally get
\begin{eqnarray}
&&\xi=-\sum_A\frac{D-q_A-3}{\Delta_A}\ln H_A,\nonumber\\
&&\eta=\sum_A\frac{q_A+1}{\Delta_A}\ln H_A,\nonumber\\
&&\zeta_\alpha=-\sum_A\frac{\delta_A}{\Delta_A}\ln H_A.
\label{eqn:BPSconditions}
\end{eqnarray}

The conditions given by Eq.~(\ref{eqn:BPSconditions})
correspond to the BPS conditions necessary for supersymmetry.
The next question is how much supersymmetry remains.
Since this depends on what and how many branes are included,
we concentrate here on the intersecting N-branes solution
given by Eq.~(\ref{sol:cross_nm2_nm5}), even though
the extension to other models is simple.

In this case, by substituting Eq.~(\ref{eqn:BPSconditions})
into Eq.~(\ref{sol:cross_nm2_nm5}), the concrete solution
is obtained as 

\begin{eqnarray}
ds^2_{11}&=&H_2^{1/3}H_5^{2/3}[(H_2H_5)^{-1}2du(dv+fdu)\nonumber\\
&&\quad +(H_2H_5)^{-1}dy_2^2+H_5^{-1}(dy_3^2+\cdots+dy_6^2)
+H_2^{-1}dy_7^2+dx_1^2+dx_2^2+dx_3^2].
\label{eqn:msolutions}
\end{eqnarray}

For this solution, if and only if $\eta$ satisfies this condition, 
the Killing equation 
is satisfied, that is,
\begin{eqnarray}
\delta\psi_i&=&-\frac{1}{6}e^{\eta-\xi}\left[
\partial_v\eta\gamma_{iu}(1+\gamma_{v\alpha_2\alpha_7})
+(\partial_u-f\partial_v)\eta\gamma_{iv}
(1+\gamma_{u\alpha_2\alpha_7})\right.\nonumber\\
&&\left.+2\partial_v\eta\gamma_{iu}(1-\gamma_{v\alpha_{2}
\cdots\alpha_6})+2(\partial_u-f\partial_v)\eta\gamma_{iv}(1
-\gamma_{u\alpha_{2}\cdots\alpha_6})\right]\zeta=0.
\end{eqnarray}
Thus in order to remain supersymmetrical,
it seems that there are four conditions. However, 
it can be shown that among them, the following three are independent,
\begin{eqnarray}
(1+\gamma_{v\alpha_2\alpha_7})\zeta=0,\quad 
(1+\gamma_{u\alpha_2\alpha_7})\zeta=0,\quad
(1-\gamma_{v\alpha_{2}\cdots\alpha_6})\zeta=0.
\end{eqnarray}

Since $\gamma$'s have eigenvalues $\pm1$ with the same numbers,
half of the supersymmetry remains from each condition.
Therefore, the Killing spinor considered here has $1/8$ supersymmetry, 
which is related to the four-dimensional $N=1$ supersymmetry 
with compactification on torus. 

It is important to discuss the possibility to generate other supersymmetric
solutions based on the solutions obtained above.
In general, the dimensional reduction of 11-dimensional
supergravity provides 10-dimensional theory with two supersymmetries,
that is, type IIA supergravity. 
In this theory, the gravitinos have opposite chiralities
($\gamma$ eugenvalues), i.e., it is ``nonchiral''. 
There is the other 10-dimensional theory with two supersymmetries that 
cannot be obtained by the reduction or the truncation of the
11-dimensional theory, that is, type IIB supergravity.
In this theory, the gravitinos have the same chirality,
i.e., it is ``chiral''. 
These describe the leading low-energy behaviors of type IIA and 
type IIB superstring theory respectively.
This fact makes the explicit formulations 
of the supergravity theories of particularly interest. 

If we compactify the $y^7$ coordinate in the solutions above,
we obtain the pp-wave solutions in type IIA supergravity theory 
in which the dilaton $\varphi$ with coupling constant 
$\epsilon a=(3-q)/2$ appears. In this case, we get
\begin{eqnarray}
\varphi=\sum_A\epsilon_Aa_A\frac{D-2}{\Delta_A}\ln H_A.
\end{eqnarray}

Since for this case the harmonic rule is satisfied naturally, 
we can make all the possible solutions 
in type II supergravity theories by making use of the 
T- and S-dual transformations as in BPS Dp-barne cases.

\section{Cosmological Solution}

In the previous section, we show new classes of the solutions 
satisfying the BPS conditions in the context of the pp-wave
background. However, at this point, even after imposing the BPS conditions,
many degrees of freedom remain to fix the solutions concerned 
with the forms
of $H_2 (u,v)$, $H_5 (u,v)$, $f(u)$. On the other hand, it seems
very important to relate the time-dependent solutions that are supposed 
to describe the early stage of our Universe and the string theory.
Therefore, in this section, by choosing the forms of the functions 
appropriately, we give examples corresponding to the inflation
in our Universe, even though we do not mention how they are chosen
in detail.

After compactifying the $y_\alpha$ $(\alpha = 2, \cdots, p)$
coordinates and conformally transforming into the (d+1)-
dimensional Einstein frame, the original Einstein-Hilbert action
in the D-dimensional theory can be given by
\begin{eqnarray}
\sqrt{-g_D}R_D \propto \Omega^{d+1}\sqrt{-g_{d+1}}
\prod_A H_A^{-q_A
\frac{D-q_A-3}{\Delta_A}
+(p-q_A-1)\frac{q_A+1}{\Delta_A}}
\Omega^{-2}R_{d+1},
\end{eqnarray}
where $g_{d+1}$ and $R_{d+1}$ are the determinant and the scalar curvature
with respect to the (d+1)-dimensional metric in the Einstein frame 
$g^{(d+1)}_{\mu\nu}$. $\Omega$ is the conformal factor which relates 
$g^{(d+1)}_{\mu\nu}$ and the metric directly obtained by the compactification
$g^{\ast(d+1)}_{\mu\nu}$ as 
$g^{\ast(d+1)}_{\mu\nu} = \Omega^2 g^{(d+1)}_{\mu\nu}$. 
We omit the term from the volume of the compactified 
$y_\alpha$ coordinates.

Since in the (d+1)-dimensional Einstein frame, this term
should also scale as $\sqrt{-g_{d+1}} R_{d+1}$
$\Omega$ can be determined as 
\begin{eqnarray}
\Omega=\prod_A H_A
^{-\frac{D-q_A-3-(q_A+1)(d-2)}{(d-1)\Delta_A}}.
\end{eqnarray}

Therefore the (d+1)-dimensional metric after 
compactifying the $y_\alpha$ coordinates can be written as 
\begin{eqnarray}
&&ds^2=\Xi^{d-2}2du(dv+fdu)+\Xi^{-1}\sum_{i=1}^{d-1}dx_i^2,\nonumber\\
&&\Xi=\prod_A H_A^{-\frac{2(D-2)}{(d-1)\Delta_A}}.
\end{eqnarray}

Furthermore, in order to obtain the cosmological solutions,
that is, time-dependent ones, we now compactify the
$y_1 = (u+v)/ \sqrt{2}$ coordinates on $S^1$, by which the fucntions 
$H_2$, $H_5$, and $f$ come to depend only on $t$,
and the resulting d-dimensional metric can be written by
\begin{eqnarray}
ds_d^2&=&-\Upsilon^{d-3}dt^2+\Upsilon^{-1}
\sum_{i=1}^{d-1}dx_i^2,\nonumber\\
\Upsilon&=&(1+f)^{-\frac{1}{d-2}}\prod_A
H_A^{-\frac{2(D-2)}{(d-2)\Delta_A}}.
\end{eqnarray}

In the four-dimensional case, in which we consider the intersecting
NM2-brane and NM5-brane especially, the metric can be written as 
\begin{eqnarray}
ds^2&=&-((1+f)H_2H_5)^{-1/2} dt^2+((1+f)H_2H_5)^{1/2}
\sum_{i=1}^3dx_i^2\nonumber\\
&=&-d\tau^2+a(\tau)^2\sum_{i=1}^3dx_i^2,
\label{sol:cosmological}
\end{eqnarray}
where $\tau$ is the cosmic time and $a(\tau)=dt/d\tau$
is the scale factor of our Universe. 
In this model, the evolution of the scale factor depends on the
form of $H_2 (t)$, $H_5 (t)$, and $f(t)$. One interesting 
question about the early stage of the Universe is whether
the inflationary solutions are consistent with supersymmetry.
We find if $((1+f) H_2 H_5) \propto t^4$, the exponentialy
expanding universe $a(\tau)\propto e^{\tau}$ is realized.
One example satisying the BPS conditions is 
$H_2$, $H_5$, $f$ $\propto t^{4/3}$.
We can also show if $((1+f) H_2 H_5) \propto t^{4(p-1)/p}$,
the power law inflationary solutions in which $a(\tau)\propto \tau^{p-1}$
are obtained. It can be easily seen that a solution such as
$H_2$, $H_5$, $f$ $\propto t^{4(p-1)/3p}$ satisfies the BPS conditions.

In the above, we construct the inflationary solution from the
viewpoint of our four-dimensional spacetime by compactifying
the $y_1$ coordinate first. However, we can show the
compactification of that coordinate realizes dynamically
for the examples considered above.
A five-dimensional metric with $y_1$ direction is given by 
\begin{eqnarray}
ds^2&=&(H_2H_5)^{-2/3}2du(dv+fdu)+(H_2H_5)^{1/3}
\sum_{i=1}^{3}dx_i^2\nonumber\\
&=&\Omega^2(-d\tau^2+a(\tau)^2\sum_{i=1}^3dx_i^2
+b(\tau)^2dy^2),
\end{eqnarray}
where 
\begin{eqnarray}
&&dy=dy_1-\frac{f}{1+f}dt, \quad b(\tau,y_1)^2=\sqrt{\frac{(1+f)^3}{H_2H_5}}, 
\\&&\Omega^2=(1+f)^{-1/2}(H_2H_5)^{-1/6}.
\end{eqnarray}
Even though, strictly speaking, the $y$ coordinate does not agree with
$y_1$, it plays the same role in the limit of $f \to 0$, whose scale
factor evolves as $a/b\propto \left(H_2H_5/(1+f)\right)^{1/2}$.
Therefore, if we choose the examples in which all functions have
the same contributions as mentioned above,
for the exponentialy inflation case we find $a/b\propto e^{2\tau/3}$, and 
for the power-law inflation case we find $a/b\propto \tau^{\frac{2(p-1)}{3}}$,
which means that for these cases, the compactification of the $y$ coordinate
happens dynamically.

It is worthwhile to relate the above cosmological solution to 
other solutions based on other types of string theory.
When it comes to $y_2, \ldots, y_7$ coordinates, which are compactified
in the above analysis, following a similar idea, it can be shown 
that the volume element of the $y_2$ coordinate shrinks more rapidly
than that of $y_3, \ldots, y_7$ coordinates.
If we compactify only $y_2$ coordinate in 11-dimensional M-theory, 
we can find the $D2$ and  $NS5$-brane's bound state in type IIA string theory. 
 Furthermore, using the T-duality transformation on the $y^3$ direction, 
 the $D3$- and $NS5$-brane's bound state is obtained. 
  It is shown that the $D3$-brane is rolling in``throat geometry" 
  on the $NS5$ background \cite{Nakayama}, and the corresponding
  cosmological solutions are already provided by \cite{Yavartanoo},
 which is related to the rolling tachyon given by \cite{Sen1},
 from the point of view of the string theory.

\section{Conclusion}

In this paper we have examined a system where $D=(d+p)$- 
dimensional gravity is coupled to a scalar field and 
arbitrary rank form gauge fields. 
The corresponding action is so general that it can describe 
the bosonic part of $D=11$ or $D=10$ supergravities.

The ansatzs employed here 
are to take the pp-wave background for the metric and the 
$n=q+2$ gauge field that can describe electric $q$-brane 
form and magnetic $(D-p-d-2)$-brane form. 
In this set up, since they occupy the $u$ coordinate, they are
null-branes (N-branes). Since we are interested in 
time-dependent solutions, we demand all the functions
appearing depend only on the null coordinates $(u,v)$.

Under these conditions we constructed intersecting N-brane solutions 
in a pp-wave background. First we treated the bosonic part by considering
the action directly. As a result, we obtained the crossing rule of the
intersecting N-branes and the harmonic function rule 
given by Eq.~(\ref{eqn:crossing}) and Eq.~(\ref{eqn:harmonicrule}).
We show the crossing rule is the same as that for \cite{Ohta:2003zh}
in the previous work in the context of N-brane solutions.
This also agrees with the S-brane cases
discussed in \cite{Deger:2002ie, Ohta:2003uw}. 
It can also be shown that for the solutions 
under the above ansatzs, the conditions are
given by Eq.~(\ref{eqn:difxi}), (\ref{eqn:difeta}), (\ref{eqn:difzeta}), 
and (\ref{eqn:difdilaton}). As concrete examples, we considered 
11-dimensional supergravity.
In this theory, the electric type 3-form field is related to
the NM2-brane, while the magnetic type one is 
to the NM5-brane. The solutions including a single NM2-brane,
a single NM5-brane, and the intersecting NM2-brane
and NM5-brane satisfying the crossing rule obtained above
can be expressed as Eqs.~(\ref{sol:single_nm2}), 
(\ref{sol:single_nm5}), and (\ref{sol:cross_nm2_nm5}), respectively.

The discussion about the bosonic part is quite
general, although there remain many degrees of the freedom about the
functions of the metric and the scalar field.

Next, we considered the fermionic part and the conditions under which the 
solutions saisfy supersymmetry (BPS conditions). We also constrcted
BPS pp-wave brane solutions in the context of 11-dimensional supergravity. 
In this analysis, instead of
considering the action of the fermionic part directly,
we used the supersymmetry transformation of the gravitino, the so-called
Killing equation, into which we substituted the results of 
the bosonic part. If we impose the supersymmetry, the functional forms
of the solutions are limited further given by 
Eq.~(\ref{eqn:BPSconditions}). From the Killing equation, 
how much supersymmetry remain in the resultant BPS solutions
is determined, which depends on the solutions. For example, in the
case of the intersecting N-branes solution given by 
Eq.~(\ref{sol:cross_nm2_nm5}), since three
independent conditions must be satisfied, $1/8$ supersymmetry 
remains. Even though we have limited
the 11-dimensional case only for simplicity, this argument can be 
extended for the 10-dimensional case easily.

At this point, all the remaining degrees of freedom to fix the
solutions are $H_2 (u,v)$, $H_5 (u,v)$, which are the harmonic functions
related with N-branes and $f(u)$, which appeared
in the metric ansatz. Even though we have started from 
11-dimensional theory,
by dimensional reduction and applying S and T-duality trasformations, 
we could also obtain all standard intersecting BPS brane solutions 
with pp-wave in 10-dimensional type II theories.

Finally, we applied the BPS solutions to the cosmological setting 
by compactifying the extra coordinates, even though we have not mentioned
the details of the mechanism of the compactification. 
As is shown by Eq.~(\ref{sol:cosmological}), the evolution of 
the scale factor depends on the form of $H_2 (t)$, $H_5 (t)$,
and $f(t)$, which are arbitrary even after imposing the
BPS conditions. By choosing the functions appropriately,
we obtained the exponentially expanding Universe, as well as
the power-law inflationary solutions. Since these inflationary solutions
are consistent with supersymmetry, it seems interesting, even though
the mechanism to fix the functions is unclear at this time.
Furthermore, if we concentrate on the above inflationary solutions,
even without compactifying the $y_1$ direction at first,
the compactification of the corresponding coordinate happens dynamically.

From the view point of the realistic cosmology, we cannot resist
asking whether the standard big-bang Universe is recovered.
One interesting scenario is the above inflationary solution becomes
unstable as a result of supersymmetry breaking, and the inflation
terminates. After supersymmetry breaking, since we need not take account
of the BPS conditions, all we have to consider are Eqs.~(\ref{eqn:difxi}), 
(\ref{eqn:difeta}), and (\ref{eqn:difzeta}), which are by far milder than
the BPS conditions. For example, if we choose
$\zeta_{\alpha} =$const provides $\xi=$const, $\eta=$const 
and $f/H_A=$const, the cosmic expansion law of the radiation 
dominated Universe is obtained in the limit of only $H_A = H_A (u) = f(u)$.
It is also known that supersymmetry breaking generates potential heat,
which describes the reheating process, even though we do not
mention the details, and we would like to leave them for future work.

\section*{Acknowledgments}
M.T. was supported by the 21st Century COE Program 
(Holistic Research and Education Center for Physics Self-Organization
Systems at Waseda University).
We would like to thank C. Burgess and A. W. Peet for constructive comments 
at Cosmo-04 in Toronto. 
We also thank  Kei-ichi Maeda and Takashi Torii for useful discussions.

\end{document}